\documentclass[11pt]{amsart}

\usepackage{amsmath, amsfonts, amsthm, amssymb, latexsym, graphicx, eucal, fullpage, dsfont, algorithmic}

\usepackage{color,hyperref}
\definecolor{darkblue}{rgb}{0.0,0.0,0.6}
\hypersetup{colorlinks,breaklinks,
            linkcolor=darkblue,urlcolor=darkblue,
            anchorcolor=darkblue,citecolor=darkblue}

\def\CC{\mathbb C}
\def\ZZ{\mathbb Z}
\def\RR{\mathbb R}

\newcommand{\bra}[1]{\langle #1|}
\newcommand{\ket}[1]{|#1\rangle}

\newcommand{\op}{\operatorname}

\newcommand{\Ket}[1]{\left|#1\right>}

\newcommand{\tqft}{\textbf{T}}

\numberwithin{equation}{section}

\newtheorem{intro-thm}{Theorem}
\newtheorem{intro-cor}{Corollary}
\newtheorem*{thm-restate-main}{Theorem~\ref{thm:main}}
\newtheorem*{thm-restate-func}{Theorem~\ref{thm:func}}
\newtheorem*{cor-restate-cor1}{Corollary~\ref{cor:cor1}}
\newtheorem*{cor-restate-cor2}{Corollary~\ref{cor:cor2}}
\newtheorem*{lem-star}{Lemma}
\newtheorem{dfn}{Definition}[section]
\newtheorem{thm}{Theorem}[section]

\makeindex

\begin{document}

\title{Quantum algorithms\\ for invariants of triangulated manifolds}

\author{Gorjan Alagic}
   \address{Institute for Quantum Computing, University of Waterloo}
   \email{galagic@gmail.com}

\author{Edgar A. Bering IV}

\begin{abstract}
One of the apparent advantages of quantum computers over their classical counterparts is their ability to efficiently contract tensor networks. In this article, we study some implications of this fact in the case of topological tensor networks. The graph underlying these networks is given by the triangulation of a manifold, and the structure of the tensors ensures that the overall tensor is independent of the choice of internal triangulation. This leads to quantum algorithms for additively approximating certain invariants of triangulated manifolds. We discuss the details of this construction in two specific cases. In the first case, we consider triangulated surfaces, where the triangle tensor is defined by the multiplication operator of a finite group; the resulting invariant has a simple closed-form expression involving the dimensions of the irreducible representations of the group and the Euler characteristic of the surface. In the second case, we consider triangulated 3-manifolds, where the tetrahedral tensor is defined by the so-called Fibonacci anyon model; the resulting invariant is the well-known Turaev-Viro invariant of 3-manifolds. 
\end{abstract}

\maketitle

\section{Introduction}

The 1994 discovery of polynomial-time quantum algorithms for factoring and discrete logarithm~\cite{Shor} motivated a widespread search for more examples where quantum computers outperform their classical counterparts. Quantum topology is proving to be an interesting source of results in this area. For instance, it is known that quantum computers can simulate certain topological functors~\cite{FreedmanKitaevWang} and approximate the value of the Jones Polynomial of a link~\cite{AharonovJonesLandau, AharonovArad, FreedmanLarsenWang, GarneroneMarzuoliRasetti2, ShorJordan, WocjanYard}. In this article, we describe another way of applying ideas from quantum topology to quantum algorithms: approximating invariants of manifolds by contracting tensor networks constructed from triangulations.

Recall that an $n$-dimensional manifold is a topological space which is locally homeomorphic (i.e., topologically equivalent) to $\RR^n$. We will consider the cases $n=2$ and $n=3$: surfaces and 3-manifolds. An invariant of $n$-manifolds (for some fixed $n$) associates a complex number to every $n$-manifold, such that homeomorphic manifolds are associated to the same number. A simple nontrivial manifold invariant is the \emph{genus} of surfaces, which is computed simply by counting the number of handles. If the surface is given in terms of a triangulation, then we can compute the genus by the formula (2 - \#vertices + \#edges - \#faces )/2. The invariants studied in this article are more complicated than the genus, but the idea is essentially the same: starting from a triangulation of a manifold, we apply some formulas to calculate a complex number; if these formulas satisfy certain properties, the number will be an invariant.

Our goal in the first section is to give an accessible presentation of this idea in the case of two-dimensional topological lattice field theories (or TLFTs) attached to finite groups. The idea behind two-dimensional TLFTs is based on a theorem of Pachner~\cite{Pachner}, which states that any two triangulations of the same surface differ by a finite sequence of simple combinatorial moves of two types. It follows that decorating a triangulated closed surface\footnote{Unless explicitly stated, surfaces (and more generally, manifolds) are assumed to have no boundary. Requiring that a surface be closed ensures that there are also no punctures.} with tensors which are invariant under the two Pachner moves results in a topologically invariant scalar. When the tensors are defined by the multiplicative structure of a finite group, this scalar has a simple expression in terms of the Euler characteristic of the surface and the dimensions of the irreducible representations of the group. Mednykh's formula~\cite{Mednykh} states that this scalar can also be expressed in terms of maps on the fundamental group of the surface. The quantum algorithm we present approximates this scalar by contracting the TLFT tensor network using well-known methods~\cite{AradLandau}. Although this algorithm does not provide a quantum speedup for the task of calculating surface invariants, it is nonetheless useful as a simple prototype of a much more general construction. Moreover, it is conceivable that it could provide a speedup for approximating some other quantity involving the relevant group and surface.

In the second section, we study tensor networks arising from triangulations of 3-manifolds. Just as surfaces can be triangulated into a collection of triangles glued along their edges, 3-manifolds can be triangulated into a collection of tetrahedra glued along their triangular faces. In three-dimensional TLFTs, rank-six tensors are assigned to each tetrahedron and attached to one another by means of gluing tensors associated to edges of the triangulation. As shown by Turaev and Viro \cite{TuraevViro}, we can choose these tensors so that the three-dimensional versions of the Pachner moves are satisfied; this implies that the tensor network of a closed 3-manifold will contract to a topologically invariant scalar. This scalar, known as the Turaev-Viro invariant, appears to be difficult to calculate classically. No efficient classical approximation algorithms are known, and exact evaluation is NP-hard~\cite{KirbyMelvin, KirbyMelvin2}. Approximating it for manifolds specified in terms of a so-called Heegaard splitting is a BQP-complete problem~\cite{AlagicJordanKonigReichardt}. The same problem for so-called mapping tori is DQC1-complete~\cite{AlagicJordan}. An efficient quantum approximation algorithm is known when the manifold is specified in terms of Dehn surgery~\cite{GarneroneMarzuoliRasetti}. In this article, we give an efficient quantum algorithm for approximating the Turaev-Viro invariant of a 3-manifold specified by a triangulation. This result is not a consequence of the results in~\cite{AlagicJordan, AlagicJordanKonigReichardt, GarneroneMarzuoliRasetti}, as no efficient translation from triangulations to Heegaard splittings, mapping tori, or Dehn surgeries is currently known~\cite{Thurston}. However, it is possible to efficiently translate Heegaard splittings into triangulations, as we will discuss (see, for example,~\cite{BrinkmannSchleimer}.) This translation allows us to conclude that approximating the Turaev-Viro invariant of a triangulation is BQP-hard, in an appropriate sense.

As in all of the other aforementioned results, the approximation of our algorithm is additive. The approximation scale depends on the quality of the input triangulation as well as the order of tensor contraction selected by the algorithm designer. In the worst case, the approximation scale is exponential in the size of the input triangulation. On the other hand, we will show that every 3-manifold admits triangulations which are efficient in a certain sense, and where a large portion of the tensor network can be contracted unitarily, i.e., in a way that does not worsen the approximation scale. 

\section{Tensor networks and 2D topological lattice field theories}

\subsection{Tensors and tensor networks}

We begin with a brief review of tensors and tensor networks; for a gentler introduction to tensors, see~\cite{Tensors}. Throughout this section, $V$ will denote a fixed finite-dimensional vector space over $\CC$. Recall that an $(m, n)$\emph{-tensor} (over $V$) is an element of $V^{\otimes m} \otimes (V^*)^{\otimes n}$. Such a tensor is said to have \emph{contravariant} rank $m$, \emph{covariant} rank $n$ and total rank $m+n$; it can be denoted by
$$
M^{i_1i_2\cdots i_m}_{j_1j_2\cdots j_n}~,
$$
where each copy of $V$ or $V^*$ is associated with a particular index. Contravariant indices are written in superscript, while covariant indices are written in subscript. One may think of tensors in a basis-dependent way by choosing an orthonormal basis for $V$. From this point of view, a tensor is simply an indexed array of scalars. The above tensor could then be written in Dirac notation as
$$
\sum M^{i_1i_2\cdots i_m}_{j_1j_2\cdots j_n} |i_1,i_2,\dots,i_m\rangle \langle j_1,j_2, \dots, j_n|~,
$$
where the sum varies each index over the entire basis of $V$. We form the product of two tensors by composing them as multilinear maps; the use of index notation is crucial here, as two multilinear maps can be composed in many ways. For instance, the product of a vector $v$ with a dual vector $w$ could either be an 
inner product $v^iw_i \in \CC$ (if we choose to ``act on the same index'') or an outer product $v^iw_j \in V \otimes V^*$ (if we ``act on different indices''). Similarly, a matrix $A$ acting on a vector $v$ is denoted by $A^i_jv^j$, the product of $A$ with another matrix $B$ is denoted by $A_i^jB_j^k$, and the trace of $A$ is simply $A_i^i$. Notice that from the basis-dependent point of view, summation over repeated indices is implicit. In the language of tensor analysis, we say that these repeated indices are \emph{contracted}. Clearly, the product of a series of tensors is itself a tensor, and its rank is determined by the indices appearing only once in the entire product. In particular, if all the indices are contracted (as in $A_i^i$), then the resulting tensor is simply a scalar in $\CC$.

\begin{figure}[h]
\begin{center}
\includegraphics{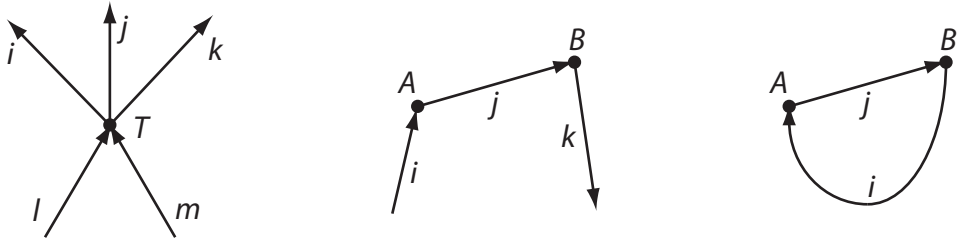}
\end{center}
\caption{Examples of simple tensor networks (from left to right): a $(3,2)$ tensor $T^{ijk}_{lm}$, the matrix product $A_i^jB_j^k$, and the matrix inner product $A_i^jB_j^i$.}
\label{fig:tensor-net}
\end{figure}

\emph{Tensor networks} provide a natural way of depicting and manipulating products of many tensors. Imagine an $(m,n)$-tensor as a widget consisting of one vertex and $m+n$ directed edges, each one attached to the vertex at one end. Each edge is labeled by an index of the tensor, and is directed out if and only if the corresponding index is contravariant. A product of two tensors then corresponds to attaching the two widgets together by appropriately joining the edges corresponding to indices shared by the two tensors. A few basic examples of tensor networks are depicted in Figure \ref{fig:tensor-net}.
If all of the edges in a tensor network are attached to vertices on both ends, then we say that the network is \emph{closed}.
\begin{dfn}
A closed tensor network $(\mathcal G, M)$ is a finite directed multigraph $\mathcal G$ equipped with tensors $M(v)$, one for each vertex $v$, so that the edges from $v$ to $w$ in $\mathcal G$ are in one-to-one correspondence with indices which are both contravariant in $M(v)$ and covariant in $M(w)$.
\end{dfn}
\noindent A closed tensor network is simply another representation of the product of the tensors $M(v)$, and is thus itself a tensor which we will denote by $T(\mathcal G, M)$. Since each edge is attached to a vertex on either end, each index appears twice in $T(\mathcal G, M)$, which is thus an element of $\CC$. This value has a simple expression in any given orthonormal basis of $V$, as follows. A \emph{labeling} of $\mathcal G$ is an assignment of a basis element to every edge of $\mathcal G$. Such a labeling $\ell$ identifies a particular entry $M(v)_\ell \in \CC$ of each tensor $M(v)$. It's then easy to check (e.g., by writing tensors in Dirac notation as above) that tensor contraction corresponds to taking sums of products of these scalars. In particular,
\begin{equation}\label{eq:state-sum}
T(\mathcal G, M) = \sum_{\text{labelings } \ell} \prod_{\text{ vertices }v} M(v)_\ell~.
\end{equation}

We now briefly summarize Arad and Landau's quantum algorithm for producing an additive approximation of the above number~\cite{AradLandau}. The algorithm involves a choice of ordering $\{M_j\}_{j=1}^n$ of the vertices of the network, according to which the tensors are contracted one-by-one. At the $j$th stage of the algorithm, we view the network as being divided into three pieces:
\begin{itemize}
\item the network $N(1,j-1)$ formed by the tensors $M_1, \dots, M_{j-1}$;
\item the network $N(j+1,n)$ formed by the tensors $M_{j+1}, \dots, M_n$;
\item the tensor $M_j$, connected to $N(1,j-1)$ by some $k$ indices and to $N(j+1,n)$ by some $l$ indices.
\end{itemize}
The algorithm maintains a register which, at this $j$th stage, contains an approximation of $N(1,j-1)$, viewed as a rank $k$ tensor\footnote{If we ignore the distinction between contravariant and covariant indices, any rank $k$ tensor can be viewed as an element of $V^{\otimes k}$ and can thus be stored (up to normalization issues) in a quantum register.}. We view $M_j$ as a map 
\begin{equation}\label{tensor-contraction-operator}
M_j: V^{\otimes k} \rightarrow V^{\otimes l}~.
\end{equation}
We would like to implement this map on our register, so that it then contains an approximation of the rank $l$ tensor $N(1,j)$. To achieve this, we first convert $M_j$ into a square matrix by appropriately adding input registers (if $k < l$) or output registers (if $k > l$). We then classically compute the singular value decomposition of this square matrix. Using a clever trick (see Lemma 3.3 of~\cite{AradLandau}) we can implement the diagonal matrix of singular values as a unitary operator, at an approximation scale cost of the operator norm $\|M_j\|$ of $M_j$, viewed in the form \eqref{tensor-contraction-operator}. At the final stage of the algorithm, having contracted all of the tensors, the algorithm approximates the value by means of the Hadamard test. The following is a restatement of this result (Theorem 3.4 of \cite{AradLandau}.)
\begin{thm}\label{thm:arad-landau}
Let $(\mathcal G, M)$ be a closed tensor network where every tensor has degree no more than $d$ and is defined over a fixed vector space $V$. Let $\{M_j\}_{j=1}^n$ be an ordering of the tensors of $(\mathcal G, M)$, and let $\Delta = \prod_{j=1}^n \|M_j\|$ be the product of the norms of the resulting contraction operators. Given any $\epsilon > 0$, there exists a quantum algorithm that runs in time $\emph{poly}((\dim V)^d)n/\epsilon^2$ and outputs a complex number $x$ such that
$$
\emph{Pr}\left(|T(\mathcal G, M) - x| \geq \epsilon \Delta\right) \leq \frac{1}{4}.
$$
\end{thm}
\noindent In the remainder of the paper, we will discuss how this algorithm can be applied to approximate certain invariants of triangulated manifolds.\\

\subsection{Surface invariants from 2D topological lattice field theories}\label{sec:tlft}

A theorem of Rad\'o from 1925 states that every orientable surface admits a triangulation~\cite{Rado}. A triangulation is a purely combinatorial object, consisting of a list of triangles along with ``gluing'' rules which tell us how to attach the triangles to each other.\footnote{Note that this is quite different from the notion of ``triangulation'' used in computer graphics. Our topological notion of triangulation does not describe a way of embedding the triangles in $\RR^3$, and we allow isotopies for free (i.e, triangles are not required to be flat or rigid.) Moreover, we tend to prefer coarse triangulations, since they are simpler combinatorial descriptions of the same topological object.} Each triangle inherits an orientation from the surface, and each of its edges are glued to exactly one other triangle. The simplest triangulation of the torus, for instance, is formed by making a square sheet out of two identical right-angled triangles, and then gluing the opposite edges of the sheet. We point out two ways to change this triangulation without changing the topological nature of the torus. First, we could replace the two triangles by another pair of triangles, glued along the other diagonal of the square sheet. Second, we could have subdivided any one triangle into three by choosing a point somewhere in its interior and adding three new edges. These two moves are called the 2-2 move, and the 1-3 move, respectively. It is clear that performing any of these moves (or their inverses) does not change the topology of the underlying surface, since it amounts to replacing a disk with a disk. A beautiful theorem of Pachner~\cite{Pachner} states that any two triangulations of the \begin{figure}[h]
\begin{center}
\includegraphics{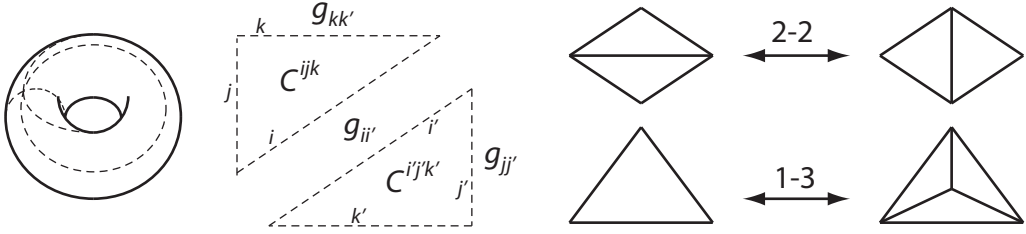}
\end{center}
\caption{Left: a simple triangulation of the torus, and the corresponding tensor network. Right: the 2-2 and 1-3 Pachner moves on triangulations.}
\label{fig:triangulation}
\end{figure}
same surface differ by only a finite sequence of these moves. Pachner's theorem suggests that we might find topological invariants by looking for objects which are unchanged by the 2-2 and 1-3 moves. For instance, it's easy to check that the quantity \#vertices - \#edges + \#faces (known as the \emph{Euler characteristic}) is invariant under these moves; in fact, one can show that any linear combination of the number of vertices, edges and faces which is invariant under the Pachner moves is a scalar multiple of the Euler characteristic~\cite{Roberts}. The Euler characteristic of a surface of genus (i.e., number of ``handles'') $g$ is equal to $2-2g$. Since the genus is a complete invariant of closed orientable surfaces, so is the Euler characteristic.

Of course, we can imagine defining more complicated algebraic objects which are invariant under the Pachner moves; topological lattice field theories (or TLFTs) are based on this idea. A TLFT is defined by a complex vector space $V$ and tensors $C^{ijk}$ and $g_{ij}$ (over $V$) which satisfy certain tensor equations. Given this data, we can associate a tensor to every triangulated surface, as follows. Separate out each triangle so that the number of edges is now exactly three times the number of triangles. Label each edge of each triangle with an index, i.e., a copy of $V$. For each triangle, assign the tensor $C^{abc}$, where $a$, $b$, and $c$ are the indices of its edges. For each pair of edges $s$ and $t$ that should be glued back together to form the original triangulation, assign the tensor $g_{st}$. The result is a tensor network - the torus case is shown in Figure \ref{fig:triangulation}. If an index appears on an interior edge of the surface, then it is contracted, as it is covariant in a gluing tensor and contravariant in a triangle tensor. The rank of the entire tensor is thus equal to the number of edges on the boundary of the initial triangulation. In particular, if the surface has no boundary, the result is a scalar. We ensure that this tensor is topologically invariant by requiring that $C^{ijk}$ is invariant under cyclic permutations of its indices (as these are the allowed orientation-preserving transformations of a triangle), and that both $C^{ijk}$ and $g_{ij}$ satisfy the Pachner moves. For instance, the 2-2 move corresponds to the tensor equation
\begin{equation}\label{eq:associativity}
C^{abc}g_{cd}C^{def} = C^{acf}g_{cd}C^{bed}~.
\end{equation}
This motivates the following definition.
\begin{dfn}
A two-dimensional topological lattice field theory (or 2-D TLFT) is a triple consisting of a finite-dimensional complex vector space $V$, a cyclically invariant $(3,0)$-tensor $C^{ijk}$ (called the triangle tensor) and a $(0,2)$-tensor $g_{ij}$ (called the gluing tensor), both defined over $V$, which satisfy the 2-2 and 1-3 Pachner moves. 
\end{dfn}
\noindent As it turns out, 2-D TLFTs can be classified completely. Notice that the tensor $C^{ijk}g_{kl}$ is a multiplication map from $V \otimes V$ to $V$, equipping $V$ with an algebra structure. From this point of view, equation \eqref{eq:associativity} is exactly the associativity property of the algebra product. One can show that invariance under the 1-3 move is equivalent to semisimplicity of $V$. We thus have that two-dimensional TLFTs are in one-to-one correspondence with associative, semisimple algebras~\cite{FukumaHosonoKawai}. A natural family of such algebras are the group algebras of finite groups. 
   
Given a finite group $G$, the group algebra $\CC G$ is the vector space of maps from $G$ to $\CC$, with algebra product given by extending the group product by linearity. The natural action $[x \cdot f](y) = f(x^{-1}y)$ of $G$ on elements of $\CC G$ turns the group algebra into a representation of $G$, called the \emph{right regular representation}. It's easy to check that the character of this representation, denoted $\chi_{\text{reg}}$, satisfies
$$
\chi_{\text{reg}}(x) =
\begin{cases}
|G| & \text{ if } x = 1\\
0 & \text{ otherwise.}
\end{cases}
$$
The tensors of the 2-D TLFT associated to a finite group $G$ are then defined by setting
$$
C^{ijk} = \frac{\chi_\text{reg}(ijk)}{|G|^{3/2}} \qquad \text{and} \qquad g_{ij} = \frac{\chi_\text{reg}(ij)}{|G|}~.
$$
The tensor $C^{ijk}$ is clearly cyclically invariant, and the 2-2 move is satisfied due to the associativity of the group product. It is straightforward to check that, with the above scaling, these tensors are also invariant under the 1-3 move. 

We can thus associate a number $M(G,S) \in \CC$ to every pair $(G, S)$ where $G$ is a finite group and $S$ is a triangulated closed orientable surface, such that $M(G,S) = M(G, f(S'))$ for any choice of triangulation $S'$ of $S$ and homeomorphism $f$. An elegant derivation of this invariant is described in~\cite{Snyder}; it is given by
\begin{equation}\label{eq:tqft-invariant}
M(G, S) \triangleq \sum_{\rho \in \hat G} \dim(\rho)^{\chi(S)}~,
\end{equation}
where $\chi(S)$ is the Euler characteristic of $S$ and $\hat G$ is the set of irreducible representations of $G$. Mednykh's Formula~\cite{Mednykh} states that
$$
\sum_{\rho \in \hat G} \dim(\rho)^{\chi(S)} = |G|^{\chi(S)-1}|\text{Hom}(\pi_1(S), G)|~,
$$
where $\pi_1(S)$ denotes the fundamental group of $S$. A beautiful recent proof of this formula, due to Snyder, proceeds by computing the value of the TLFT invariant in the group basis (resulting in the right-hand side) and in the Fourier basis (resulting in the left-hand side)~\cite{Snyder}. 

\subsection{Quantum algorithm for approximating the 2D TLFT invariant}

\noindent The previous section described how to create a tensor network associated to a finite group $G$ and a triangulated surface $S$. To apply Theorem \ref{thm:arad-landau} to this case and produce an algorithm for approximating the invariant $M(G, S)$, we need to determine the operator norms of contracting $C^{ijk}$ and $g_{ij}$ in various ways, and discuss the choice of ordering.

To compute the relevant operator norms, it is useful to write our tensors in Dirac notation. In what follows, all sums are taken over the group $G$. The triangle tensor, viewed as a $(3,0)$ tensor, is equal to
$$
\frac{1}{\sqrt{|G|}} \sum_{abc=1} \ket a \ket b \ket c~.
$$
If we write it instead as a $(2,1)$ tensor and a $(1,2)$ tensor, then we get
$$
\frac{1}{\sqrt{|G|}} \sum_{abc=1} \ket a \ket b \bra c \qquad \text{and} \qquad \frac{1}{\sqrt{|G|}} \sum_{abc=1} \ket a \bra b \bra c~,
$$
respectively. It is straightforward to check that the $(3,0)$ and $(0,3)$ cases each have operator norm $\sqrt{|G|}$ while the $(2,1)$ and $(1,2)$ cases have operator norm $1$. The gluing operator, viewed as a $(0,2)$ tensor and a $(1,1)$ tensor, is equal to
$$
\sum_{ab=1} \ket a \ket b \qquad \text{and} \qquad \sum_{ab=1} \ket a \bra b~,
$$ 
respectively. It is again straightforward to check that the operator norms are $\sqrt{|G|}$ and $1$, respectively. 

We note that the operators defined above have a rather special structure. Theorem \ref{thm:arad-landau} is quite general and assumes nothing about the operators $M_j$, implementing them in time polynomial in the dimension, which contributes a factor of $\text{poly}(m^d)$ to the running time. In our case, for choices of $G$ where multiplication can be performed efficiently, the $(3,0)$ and $(0,3)$ tensors (scaled appropriately) can be implemented in polylog$(|G|)$ time by forming a uniform superposition over $G \times G$ and then applying the controlled multiplication operator $\ket{a}\ket{b}\ket{0} \rightarrow \ket{a}\ket{b}\ket{b^{-1}a^{-1}}$. The $(2,1)$ and $(1,2)$ tensors can be implemented efficiently in a similar manner.

The total approximation scale of our algorithm is the product of the operator norms of the various tensors, contracted in the order which we choose. This scale will increase by a factor of $\sqrt{|G|}$ whenever a triangle tensor is contracted as a $(0,3)$ or a $(3,0)$ tensor, or when a gluing tensor is contracted as a $(0,2)$ or a $(2,0)$ tensor. Now, let $S$ be a triangulated orientable surface equipped with an ordering $\{S_j\}$ of the set of all edges and triangles in the triangulation. We will call an edge $e$ a \emph{cap} if the two triangles surrounding $e$ in the triangulation are either both before $e$, or both after $e$ in the ordering $\{S_j\}$. Similarly, a triangle $t$ is a cap if the three edges surrounding $t$ in the triangulation are either all three before $t$, or all three after $t$ in the ordering $\{S_j\}$. We denote the total number of caps by $k(S)$. It is clear that $k(S)$ also denotes the number of $\sqrt{|G|}$ operator norm factors resulting from contracting the tensor network of $S$ in the order $\{S_j\}$.

By the discussion above, we can now specialize Theorem \ref{thm:arad-landau} to the case of approximating the TLFT invariant $M(G, S)$ attached to an orientable surface $S$ and a finite group $G$. 
\begin{thm}\label{thm:2Dalgorithm}
Let $G$ be a finite group and let $S$ be a compact orientable surface equipped with a triangulation of size $n$ and an ordering of the set of all edges and triangles; let
$$
M(G,S) = \sum_{\rho \in \hat G} d_\rho^{\chi(S)} = |G|^{\chi(S)-1}|\emph{Hom}(\pi_1(S), G)|~.
$$
Then for every $\epsilon > 0$, there exists a quantum algorithm which runs in time $\text{polylog}(|G|^3)n/\epsilon^2$ and outputs a complex number $x$ such that
$\emph{Pr}(|M(G,S) - x | \geq \epsilon|G|^{k(S)/2}) \leq 1/4.$
\end{thm}
\noindent This algorithm is a simple and natural example of applying tensor network contraction to computing topological invariants. However, as we now discuss, due to issues of approximation scale and the known efficient classical algorithms for computing some of the quantities involved in $M(G,S)$, Theorem \ref{thm:2Dalgorithm} is unlikely to provide any quantum speedup. 

\begin{figure}[h]
\begin{center}
\includegraphics{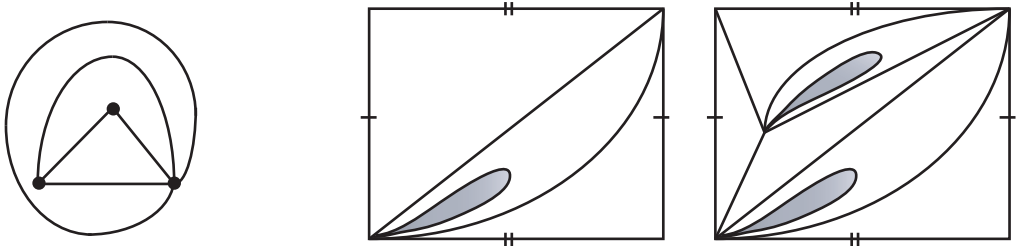}
\end{center}
\caption{Left: inserting this triangulation of the disk will force a cap, worsening the approximation scale by a factor of $\sqrt{|G|}$. Right: triangulations of the torus with one and two deleted disks(shaded); gluing $g$ of these together along the boundaries in the appropriate way results in a torus of genus $g$, having only two caps and a number of triangles which is linear in $g$.}
\label{fig:baddisk}
\end{figure}

First, it is clear that $k(S)$ is always at least two, since the first and last elements of $\{S_j\}$ are always caps. Moreover, $k(S)$ could be as bad as linear in $n$; any given triangulation can be worsened by inserting the triangulation of the disk shown in Figure \ref{fig:baddisk}. If we assume that each edge of the disk is contracted in a $(1,1)$ fashion, then it follows that one of the triangles must be contracted $(0,3)$ or $(3,0)$. This is simply because the disk has only one exterior edge, and thus only two ways (up to direction) of contracting the interior - both of which result in a triangle cap. Inserting such a disk will thus increase $k(S)$ by one and the size of the triangulation by three. On the other hand, it is easy to come up with triangulations of genus $g$ surfaces with $k(S)=2$ (see Figure \ref{fig:baddisk}); this leads to the best possible approximation scale of $|G|$. 

There are several quantities in Mednykh's formula that we might try to approximate using the above algorithm. First, we could imagine choosing a group for which the dimensions of the irreducible representations are known, and then using the above algorithm to approximate $\chi(S)$. However, since $\chi(S) = 2 - 2 g(S)$ (where $g(S)$ is the genus of $S$), this approximation scale is exponentially large in the genus, even in the ideal case where some $\rho \in \hat G$ is of dimension nearly $\sqrt{|G|}$. Moreover, $\chi(S)$ is trivial to compute exactly using a classical algorithm.

We might also imagine a scenario (perhaps a black-box problem) in which we know how to implement the tensors $C^{ijk}$ and $g_{ij}$ for some finite group $G$, but we do not know the dimensions of the irreducible representations of $G$. We could then choose specific inputs of $S$ to try and gain some representation-theoretic information about $G$. For instance, if $g(S) = 0$ then $M(G,S) = |G|$ and if $g(S) = 1$ then $M(G, S) = |\hat G|$, and so on. Unfortunately, even if we choose the triangulations ourselves, the approximation scale is still never better than $|G|$. 

Finally, we remark that a simple classical probabilistic algorithm for approximating $M(G, S)$ has similar performance to the quantum one from Theorem \ref{thm:2Dalgorithm}. In the case of $M(G, S)$, a term in the sum \eqref{eq:state-sum} corresponds to an assignment of group elements to each edge of every triangle in the triangulation. An assignment is \emph{valid} if the cycle of edges around each triangle multiplies to the identity and if every glued pair of edges multiply to the identity. Only the valid labelings contribute non-zero terms, and each of those terms has the same value: $|G|^{\#\text{triangles}-\#\text{edges}}$. Thus 
$$
M(G,S) = |G|^{\#\text{triangles}-\#\text{edges}}\#\{\text{valid labelings}\}.
$$
This suggests a simple-minded classical algorithm: generate labelings randomly and count the number of valid ones among them. A simple analysis shows that this algorithm has similar performance to the quantum one. An important distinction is that the quantum algorithm involves a choice of ordering of the triangulation.

\section{Three-dimensional topological lattice field theories}

\subsection{The Turaev-Viro invariant of 3-manifolds}\label{sec:TV}
\begin{figure}[h]
\begin{center}
\includegraphics[width=0.9\textwidth]{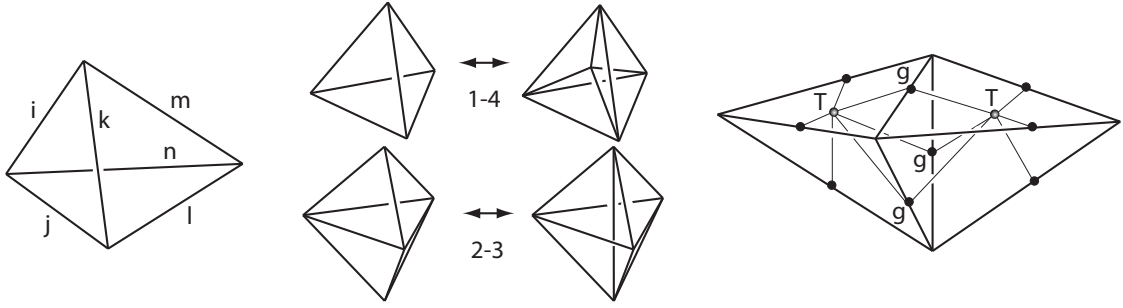}
\end{center}
\caption{Left: a labeled tetrahedron. Middle: the three-dimensional Pachner moves. Right: the tensor network associated to two tetrahedra glued together along a common triangle; the symbol tensor is denoted by $T$, and the gluing tensors are denoted by $g$.}
\label{fig:tetnet}
\end{figure}

In 1992, Turaev and Viro introduced a new invariant of triangulated 3-manifolds~\cite{TuraevViro}. Their construction, which we now outline, is a three-dimensional analogue of the two-dimensional TLFT discussed in the previous section. We begin by choosing a finite index set $I$, a function $d:I \rightarrow \CC$, and a set $J  \subset I^3$ of triples from $I$. These pieces of data are referred to by various names in the literature: the elements of $I$ are called \emph{labels} or \emph{particle types}; the value $d_i$ is called the \emph{quantum dimension} of particle type $i$; the elements of $J$ are called \emph{admissible triples} or \emph{fusion rules}. The final piece of data we need is a function which maps each six-tuple $(i,j,k,l,m,n)$ of elements from $I$ to a single complex number:
$$
\begin{vmatrix} i & j & k \\ l & m & n \end{vmatrix} \in \CC~;
$$
we will refer to this as the \emph{symbol tensor}. The symbol tensor corresponds to a tetrahedron whose six edges are labeled by $(i,j,k,l,m,n)$ as shown in Figure \ref{fig:tetnet}. It will take the value zero unless all four of its triangular faces $(i,j,k),(k,l,m),(m,n,i),(j,l,n)$ are admissible; in this 
case, we will say that the six-tuple $(i,j,k,l,m,n)$ is also admissible. The symbol tensor must be invariant under the symmetries of the tetrahedron:
\begin{equation}\label{eq:tet-symmetries}
\begin{vmatrix} i & j & k \\ l & m & n \end{vmatrix} = 
\begin{vmatrix} j & i & k \\ m & l & n \end{vmatrix} = 
\begin{vmatrix} i & k & j \\ l & n & m \end{vmatrix} = 
\begin{vmatrix} i & m & n \\ l & j & k \end{vmatrix} = 
\begin{vmatrix} l & m & k \\ i & j & n \end{vmatrix} = 
\begin{vmatrix} l & j & n \\ i & m & k \end{vmatrix}~.
\end{equation}
Using this data, we can turn any triangulated 3-manifold $M$ without boundary into a tensor network, as illustrated in Figure~\ref{fig:tetnet}. Let $\mathcal V$, $\mathcal E$, and $\mathcal T$ denote the vertex set, edge set, and tetrahedron set of $M$, respectively. The graph underlying the tensor network will have vertex set $\mathcal E \cup \mathcal T$. Two vertices in this graph will only be connected if one of them is a tetrahedron $t \in \mathcal T$, and the other an edge $e$ of $t$. To turn this graph into a tensor network, we assign the symbol tensor (denoted $T$ in the figure) to each element of $\mathcal T$, and a gluing tensor (denoted $g$) to each element of $\mathcal E$; this gluing tensor will take the value $d_i$ whenever all of its indices have the same label $i$, and zero otherwise. The gluing tensor $g$ associated to edge $e$ will have rank equal to the number of tetrahedra incident on $e$.

Every index of the above network will appear in both a gluing tensor and in a symbol tensor, so that the overall network will evaluate to a complex number. This number can be specified as follows. A labeling of $\mathcal E$ is a map $\phi: \mathcal E \rightarrow I$ such that the labels assigned to each triangle are admissible. Such a labeling associates to each tetrahedron $t \in \mathcal T$ a six-tuple $\phi(t)$, consisting of the labels of its edges. We denote the value of the symbol tensor of such a labeled tetrahedron by $|\phi(t)| \in \CC$. As in \eqref{eq:state-sum}, the value of this tensor network is then given by
$$
\sum_{\text{labelings }\phi} ~~~~~\prod_{e \in \mathcal E} d_{\phi(e)} \prod_{t \in \mathcal T} |\phi(t)|~.
$$

We would like to use the above to produce a number which depends only on the homeomorphism type of the underlying manifold. To achieve this, we must place some additional conditions on the symbol tensors and the gluing tensors. These conditions essentially amount to guaranteeing that the resulting tensor network is invariant under the three-dimensional Pachner moves, shown in Figure~\ref{fig:tetnet}. First, for any $j_1, \dots, j_6 \in I$ such that $(j_1, j_3, j_4)$, $(j_2, j_4, j_5)$, $(j_1, j_3, j_6)$ and $(j_2, j_5, j_6)$ are admissible, we need
\begin{equation}\label{one-star}
\sum_j d_{j}d_{j_4}
\begin{vmatrix} j_2 & j_1 & j \\ j_3 & j_5 & j_4 \end{vmatrix}
\begin{vmatrix} j_3 & j_1 & j_6 \\ j_2 & j_5 & j \end{vmatrix}
= \delta_{j_4 j_6}~,
\end{equation}
where $\delta$ is the Kronecker delta. Second, for any $j_1, \dots, j_9 \in I$ such that the 6-tuples $(j_9,j_1,j_4,j_6,j_5,j_2)$ and $(j_8,j_7,j_9,j_2,j_5,j_3)$ are admissible, we need
\begin{equation}\label{two-star}
\sum_j d_j
\begin{vmatrix} j_7 & j_1 & j \\ j_6 & j_3 & j_2 \end{vmatrix}
\begin{vmatrix} j_8 & j & j_4 \\ j_6 & j_5 & j_3 \end{vmatrix}
\begin{vmatrix} j_8 & j_7 & j_9 \\ j_1 & j_4 & j \end{vmatrix}
=
\begin{vmatrix} j_9 & j_1 & j_4 \\ j_6 & j_5 & j_2 \end{vmatrix}
\begin{vmatrix} j_8 & j_7 & j_9 \\ j_2 & j_5 & j_3 \end{vmatrix}~.
\end{equation}
Third, for any $j \in I$, we need
\begin{equation}\label{three-star}
D^2 = \frac{1}{d_j} \sum_{k,l:~(j,k,l) \in J} d_k d_l~,
\end{equation}
where $D = \sqrt{\sum_{j \in I} d_j^2}$ is the so-called \emph{total quantum dimension}. This brings us to Theorem 1.3.A of~\cite{TuraevViro}; here, the ``initial data'' refers to the label set $I$, the admissible triples $J$, the dimensions $d_i$, and the symbol tensor $|\cdot|$.
\begin{thm}\label{thm:tv-invariant}
Let $M$ be a triangulated 3-manifold with vertex set $\mathcal V$, edge set $\mathcal E$, and tetrahedron set $\mathcal T$. If the initial data specified above satisfies properties \eqref{one-star}, \eqref{two-star}, and \eqref{three-star}, then
$$
\op{TV}(M) \triangleq D^{-2|\mathcal V|} \sum_{\text{labelings }\phi} ~~~~~\prod_{e \in \mathcal E} d_{\phi(e)} \prod_{t \in \mathcal T} |\phi(t)|
$$
does not depend on the triangulation of $M$.
\end{thm}
\noindent The proof is given in~\cite{TuraevViro}. We remark that Turaev and Viro give a large number of examples of initial data which can be used to produce nontrivial invariants by means of Theorem \ref{thm:tv-invariant}. These examples use the so-called quantum 6j-symbols as the symbol tensors; the initial data, and the fact that this data satisfies properties \eqref{one-star}, \eqref{two-star}, and \eqref{three-star}, come from deep facts about the representation theory of quantum groups. The simplest of these initial data sets uses only two labels (i.e., $|I|=2$) and two admissible triples. While this makes for very simple calculations, the resulting invariant has a closed form expression involving the Betti numbers of the manifold; these can be calculated exactly using standard classical algorithms~\cite{KaibelPfetsch}.

Although all of our results apply to networks constructed from quantum 6j-symbols as in Turaev and Viro's work, as well as more general models considered by Barrett and Westbury~\cite{BarrettWestbury}, we will instead write about the so-called Fibonacci model. In this model, there are two labels and three admissible triples (up to cyclic permutations). The symbol tensor is very simple to describe, and the three essential properties required by Theorem \ref{thm:tv-invariant} can be easily verified. The resulting invariant has no known closed form expression, and appears to be computationally difficult for classical computers. Indeed, in other settings it is universal for quantum computers~\cite{AlagicJordan, AlagicJordanKonigReichardt}.

\subsection{The Fibonacci Turaev-Viro invariant}

In the Fibonacci model, the initial data from the previous section is quite simple. The label set is $I = \{0,1\}$ and the dimensions are
$$
d_0 = 1~,~
d_1 = \frac{1+\sqrt{5}}{2}~\text{ and }
D^2 = \frac{5+\sqrt{5}}{2}~.
$$
All triples are admissible so long as they do not have exactly one $1$-label. Up to cyclic permutation, these are 
$$
J = \{(1,1,1),(1,1,0),(0,0,0)\}~.
$$ This means that, among all possible six-tuples, there are fifteen admissible ones. Using equalities \eqref{eq:tet-symmetries}, one can move between any pair of six-tuples that have the same number of $0$-labels. Due to this equivalence, we need only define the symbol tensor for the following cases:
$$
\left\{\begin{vmatrix} 0 & 0 & 0 \\ 0 & 0 & 0 \end{vmatrix},~
\begin{vmatrix} 0 & 0 & 0 \\ 1 & 1 & 1 \end{vmatrix},~
\begin{vmatrix} 0 & 1 & 1 \\ 0 & 1 & 1 \end{vmatrix},~
\begin{vmatrix} 0 & 1 & 1 \\ 1 & 1 & 1 \end{vmatrix},~
\begin{vmatrix} 1 & 1 & 1 \\ 1 & 1 & 1 \end{vmatrix}\right\}~.
$$
Indeed, it is given as follows:
$$
\begin{vmatrix} j_1 & j_2 & j_3 \\ j_4 & j_5 & j_6 \end{vmatrix} = 
\begin{cases}
\frac{-2}{3+\sqrt{5}} &\text{ if }\sum_{k} j_k = 6\\
\frac{2}{1+\sqrt{5}} &\text{ if }\sum_{k} j_k = 5\text{ or }4\\
\sqrt{\frac{2}{1+\sqrt{5}}} &\text{ if }\sum_{k} j_k = 3\\
1 &\text{ if }\sum_{k} j_k = 0;\\
\end{cases}
$$
recall also that the value must be $0$ if $(j_1,j_2,j_3,j_4,j_5,j_6)$ is not admissible. Given this initial data, we can verify the properties \eqref{one-star} and \eqref{two-star} directly using a computer program; the property \eqref{three-star} is quickly verified by hand. We can now specialize Theorem \ref{thm:arad-landau} to the case of approximating the Fibonacci model Turaev-Viro invariant of a triangulated 3-manifold.

\begin{thm}\label{thm:tv-algorithm}
Let $M$ be a triangulated 3-manifold with vertex set $\mathcal V$, edge set $\mathcal E$ and tetrahedron set $\mathcal T$. Let $d'$ be the maximum degree (i.e., number of attached tetrahedra) of any edge, and let $d = \max\{6, d'\}$. Let $\{M_j\}_{j=1}^n$ be an ordering of $\mathcal E \cup \mathcal T$, and let $\Delta$ be the product of the norms of the resulting contraction operators. Given any $\epsilon > 0$, there exists a quantum algorithm that runs in time $\text{poly}(2^d)n/\epsilon^2$ and outputs a complex number $x$ such that
$$
\emph{Pr}\left(|\op{TV}(M) - x| \geq \epsilon \Delta D^{-2|\mathcal V|} \right) \leq \frac{1}{4}.
$$
\end{thm}

To analyze the approximation scale of this algorithm, we need to understand the tension between $\Delta$ and $D^{-2|\mathcal V|}$. Recall that during the algorithm of Arad and Landau~\cite{AradLandau}, each time we select a new tensor $T$ to contract, we view it as a map
$$
M_T: V^{\otimes k} \rightarrow V^{\otimes l}
$$
where $k$ is the number of incoming indices, and $l$ is the number of outgoing indices. In our case, $V$ is the complex span of $I = \{0,1\}$, i.e., a single qubit. To ``contract'' $T$, we convert $M_T$ to a square matrix and then implement it approximately as a unitary operator. The cost of this operation is a multiplicative increase of $\Delta$ by a factor of $\|M_T\|$, where $\| \cdot \|$ denotes the operator norm. In our case, we can give some simple bounds on $\|M_T\|$. First, if $T$ is a symbol tensor, then
$$
\begin{vmatrix} 0 & 0 & 0 \\ 0 & 0 & 0 \end{vmatrix} = 1
$$
implies that regardless of $k$ and $l$, $T:\Ket{0}^{\otimes k} \mapsto \Ket{0}^{\otimes l}$, so that $\|T\| \geq 1$. Second, recall that gluing tensors take the value $d_i$ if all the indices are equal to $i$, and zero otherwise. Thus, if $T'$ is a gluing tensor then regardless of $k$ and $l$, $T': \Ket{1}^{\otimes k} \mapsto d_1 \Ket{1}^{\otimes l}$, so that $\|T'\| \geq |d_1| = (1+\sqrt{5})/2$.

Now suppose that we are building a triangulation by starting from a single tetrahedron and attaching tetrahedra one at a time, in the order with which the algorithm will be performed, while keeping track of the minimum possible contributions to $\Delta' = \Delta D^{-2|\mathcal V|}$. With only the initial tetrahedron and its six adjacent gluing tensors, we have
$$
\Delta'_1 = \frac{d_1^6}{D^8} = \frac{3+\sqrt{5}}{50}~.
$$
Now let us compute $\Delta' = \Delta'_{|\mathcal T|}$. Suppose that we are adding the $j$th tetrahedron by gluing some of its triangular faces to the existing triangulation. For our purposes, there are four possible ways to attach the new tetrahedron:
\begin{enumerate}
\item via one of its faces, resulting in three new edges and one new vertex;
\item via two of its faces, resulting in one new edge and no new vertices;
\item via three of its faces, resulting in no new edges or vertices;
\item via four of its faces, resulting in no new edges or vertices.
\end{enumerate}
Notice that according to the bounds given above, cases (3) and (4) could possibly result in $\Delta_j' = \Delta_{j-1}'$. However, both of these moves reduce the number of triangles in the surface triangulation by at least two, while only the (1) move allows us to increase the number of triangles, also by two. Since the complete triangulation of $M$ has no boundary triangles, the total number of (3) and (4) moves cannot exceed the total number of (1) moves plus one (due to the initial tetrahedron). We thus have to assume, for the purposes of best-case analysis of $\Delta'$, that half of the moves are of the form (1) or (2). In the case of move (1), we must pay a penalty of at least $d_1^3 D^2 \approx 1.17$ while for move (2) we pay at least $d_1 \approx 1.62$. In either case, we have $\Delta_j' \geq d_1^3 D^2 \Delta_{j-1}'$, and so
$$
\Delta D^{-2|\mathcal V|} \geq \frac{d_1^6}{D^8} (d_1^3 D^2)^{|\mathcal T|/2} = \frac{3 + \sqrt{5}}{50}(1.08)^{|\mathcal T|}.
$$
It follows that $\Delta D^{-2|\mathcal V|} = \Theta(2^{|\mathcal T|})$. A direct application of Arad and Landau's algorithm to this problem thus always results in an approximation scale which is exponential in the number of tetrahedra.

However, a more clever application of the tensor network contraction algorithm could result in a much better scale, by taking advantage of the sub-multiplicativity of the operator norm. The algorithm designer could carefully choose some adjacent tensors in the network and contract them as a single operator, with a possibly lower approximation cost than if he had contracted them one by one. 

\subsection{Efficient triangulations of 3-manifolds}

As we now show, every 3-manifold admits triangulations such that large portions of the resulting tensor network can be contracted without incurring an increase in the approximation scale. Moreover, there are families of triangulated 3-manifolds with arbitrarily many tetrahedra, edges, and vertices but constant approximation scale. These conclusions come from two observations, which we discuss in detail below: first, that performing a 2-2 move on a surface by attaching a tetrahedron is a unitary operation, and second, that we can efficiently triangulate Heegaard splittings and mapping tori by means of attaching tetrahedra to fixed triangulations of handlebodies.

\subsubsection{Attaching a tetrahedron in a 2-2 fashion is unitary}\label{sec:tet-unitary}

Recall that in Section \ref{sec:TV}, we built a tensor network out of a triangulated 3-manifold without boundary by assigning a gluing tensor with value $d_i$ to each edge and the symbol tensor to each tetrahedron, as in Figure \ref{fig:tetnet}. We can modify this construction slightly to handle 3-manifolds $M$ with nonempty boundary $\partial M$. We first assign tensors as usual to tetrahedra and edges in $M \setminus \partial M$. Then, to edges in $\partial M$ we assign a gluing tensor with value $\sqrt{d_i}$, with a number of indices equal to the number of tetrahedra adjacent to that edge, plus one. The result is a tensor with rank equal to the number of edges in the triangulation of $\partial M$. We will denote this tensor by $\tqft(M)$. We single out two important properties of this construction. The first is that if $\partial M = \emptyset$ then by definition we recover the construction presented in Section \ref{sec:TV} for 3-manifolds without boundary, and
$$
\tqft(M) = D^{2|\mathcal V|}\op{TV}(M)~.
$$
The second property is \emph{composability}: if $M$ is a triangulated 3-manifold and $N \subset M$ a triangulated surface such that cutting $M$ along $N$ results in two 3-manifolds $M_1$ and $M_2$, then $\tqft(M)$ is equal to the tensor formed by attaching $\tqft(M_1)$ and $\tqft(M_2)$ along the indices in $N$. 

\begin{figure}[h]
\begin{center}
\includegraphics[width=0.8\textwidth]{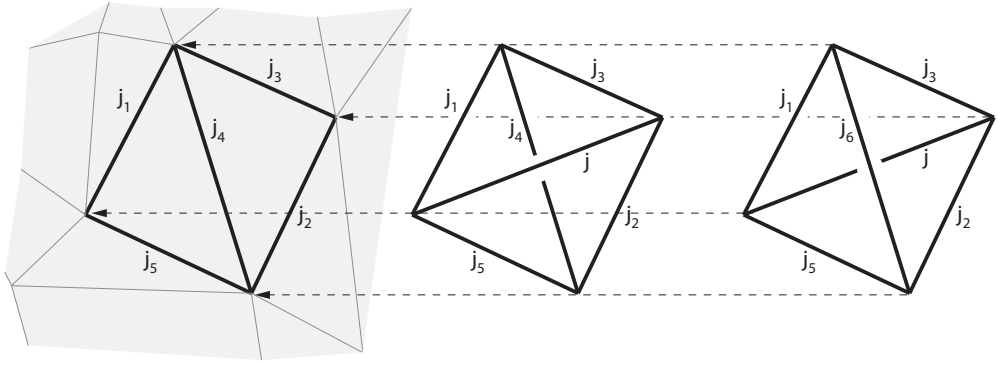}
\end{center}
\caption{We can implement 2-2 moves on a triangulated surface (left, grayed) by attaching a tetrahedron (middle). Notice that attaching a second tetrahedron (right) returns us to the original triangulation. Indeed, by \eqref{one-star}, the corresponding tensor network operator is the identity; attaching one tetrahedron is thus unitary.}
\label{fig:gluetet}
\end{figure}

Now suppose $M$ is a triangulated 3-manifold with a boundary $\partial M$ having $k$ edges. We can think of the resulting tensor as a vector $\ket{\tqft(M)} \in V^{\otimes k}$ in the complex vector space spanned by labelings of the edges of the triangulation of $\partial M$ by elements of $I$. Suppose we attach a single tetrahedron to $M$ (as in Figure \ref{fig:gluetet}), resulting in a new 3-manifold $M'$. The tensor $\ket{\tqft(M')}$ of $M'$ lives in the vector space spanned by labelings of the edges of the triangulation of $\partial M'$; this differs from the triangulation of $\partial M$ by a single flipped edge. The map that sends $\ket{\tqft(M)}$ to $\ket{\tqft(M')}$ is known as the F-move, and acts trivially on all of the indices except five:
\[
F:
\text{span}
\left\{\begin{array}{c}
\includegraphics{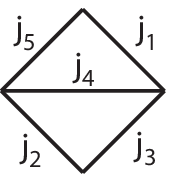}
\end{array}
: j_i \in I\right\}
\longrightarrow
\text{span}
\left\{\begin{array}{c}
\includegraphics{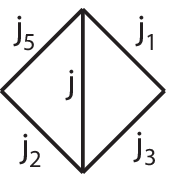}
\end{array}
: j, j_i \in I\right\}~.
\]
It is defined by
\[
\begin{array}{c}
\includegraphics{diamond1}
\end{array}
= \sum_j \sqrt{d_jd_{j_4}}
\begin{vmatrix} j_2 & j_1 & j \\ j_3 & j_5 & j_4 \end{vmatrix}
\begin{array}{c}
\includegraphics{diamond2}
\end{array}
~.
\]
Note that the matrix entries of $F^\dagger F$ (i.e., apply the F-move followed by its adjoint, as in Figure \ref{fig:gluetet}) satisfy property \eqref{one-star}:
$$
\left(F^\dagger F\right)_{j_4 j_6} 
= \sum_j F_{j j_4} \overline{F_{j_6 j}}
= \sqrt{\frac{d_{j_6}}{d_{j_4}}}\sum_j d_j d_{j_4}
\begin{vmatrix} j_2 & j_1 & j \\ j_3 & j_5 & j_4 \end{vmatrix}
\begin{vmatrix} j_3 & j_1 & j_6 \\ j_2 & j_5 & j \end{vmatrix}
= \delta_{j_4 j_6}\sqrt{\frac{d_{j_6}}{d_{j_4}}}
= \delta_{j_4 j_6}~.
$$
We conclude that $F$ is unitary.

\subsubsection{Efficient triangulations of Heegaard splittings and mapping tori}

\begin{figure}[h]
\begin{center}
\includegraphics[width=0.9\textwidth]{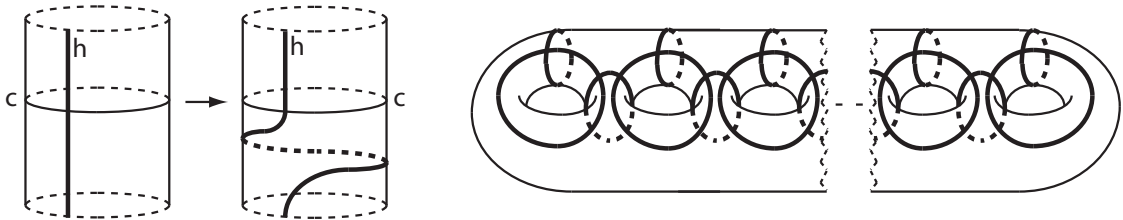}
\end{center}
\caption{Left: the Dehn twist around curve $c$ applied to a tubular neighborhood of some surface; the image of a curve $h$ under the twist is shown on the right. Right: the $3g-1$ Lickorish canonical curves (in bold) on the surface of a handlebody.}
\label{fig:dehn-twist}
\end{figure}

In the following discussion, we will denote the closed, connected, orientable surface of genus $g$ by $\Sigma_g$. Recall that such surfaces are completely characterized by their genus. The \emph{cylinder} $\Sigma_g \times [0,1]$ is a $3$-manifold with boundary consisting of two copies of $\Sigma_g$ (specifically, the bottom $\Sigma_g \times \{0\}$ and the top $\Sigma_g \times \{1\}$.) We can identify the two ends of this cylinder:
$$
\frac{\Sigma_g \times [0,1]}{(x,0) \sim (x,1)}
$$
to produce a $3$-manifold without boundary. Intuitively, the ``$\sim$'' relation means ``glue to'': each point on the bottom of the cylinder is glued to its corresponding point on the top. Now let $f$ be an orientation-preserving self-homeomorphism of $\Sigma_g$. The \emph{mapping cylinder} of $f$ is a 3-manifold with boundary, defined by
$$
M_{g, f} = \frac{(\Sigma_g \times [0,1]) \sqcup \Sigma_g}{(x, 1) \sim f(x)}~,
$$
where $\sqcup$ denotes disjoint union. This construction amounts to ``replacing'' the top of the cylinder $\Sigma_g \times [0,1]$ by gluing on the image of $\Sigma_g$ under $f$; how to glue the points together is specified by $f$. The group of isotopy classes\footnote{An isotopy between two self-homeomorphisms $f$ and $g$ of a space $X$ is a continuous deformation from $f$ to $g$ that maintains the homeomorphism property throughout. Specifically, it is a map $H:X \times [0,1] \rightarrow X$ satisfying $H(x,0)=f(x)$ and $H(x,1) = g(x)$ such that $H(\cdot,t)$ is a homeomorphism from $X$ to itself for every $t$.} of orientation-preserving self-homeomorphisms of $\Sigma_g$ is called the mapping class group, and denoted MCG$(g)$. The simplest examples are MCG$(0) \cong \{1\}$ and MCG$(1) \cong \op{SL}(2, \ZZ)$. For $g>1$, Dehn~\cite{Dehn} showed that the MCG is generated by a finite number of so-called \emph{Dehn twists} around simple closed curves on the surface. To perform a Dehn twist, take a tubular neighborhood of the curve, cut the surface along the curve, apply a $2\pi$ twist to one end of the tube, and then reglue. Lickorish~\cite{Lickorish} demonstrated a set of \emph{canonical curves} of size $3g-1$, shown in Figure \ref{fig:dehn-twist}. The Dehn twists around these curves generate all of MCG$(g)$. Any mapping cylinder $M_{g, f}$ can thus be completely specified by giving a word
$$
f = f_1f_2f_3 \cdots f_n
$$
in these standard generators of MCG$(g)$. We will assume from now on that $f$ is always given in this way.

We will consider two ways of turning a mapping cylinder $M_{g, f}$ into a closed 3-manifold without boundary. The first is the \emph{mapping torus}, produced by gluing the top of the mapping cylinder to the bottom:
$$
T_{g, f} = \frac{\Sigma_g \times [0,1]}{(x,1) \sim (f(x),0)}~.
$$
For example, choosing $g = 1$ and $f$ to be the identity map results in the three-dimensional torus. The second construction involves gluing two solid genus-$g$ handlebodies onto $M_{g, f}$, one on top and one on bottom. In this case, it might be convenient to visualize the mapping cylinder as a three-dimensional annulus, i.e., a shell in the shape of a thickened $\Sigma_g$. We can fill the interior of the shell with one handlebody, and the exterior with the other handlebody, resulting in a surfaceless 3-manifold. This manifold is called a \emph{Heegaard splitting}, and denoted by $H_{g, f}$. Every 3-manifold can be specified as a Heegaard splitting for some $g$ and $f$, but not every 3-manifold can be specified as a mapping torus~\cite{PrasolovSossinsky}.

\begin{figure}[h]
\begin{center}
\includegraphics[width=0.8\textwidth]{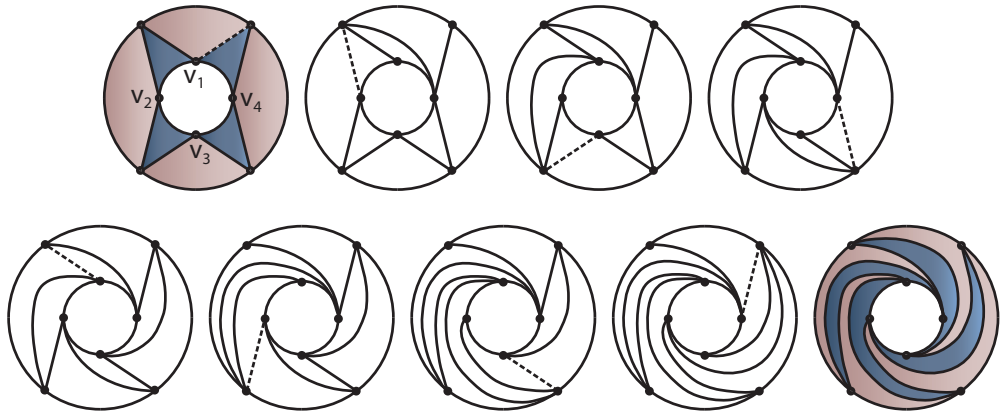}
\end{center}
\caption{We can implement a Dehn twist on a triangulated surface by means of a sequence of 2-2 moves. The above annuli represent a strip of triangles living on a surface, along a canonical curve. At each step, the dashed edge is flipped. Performing the entire sequence above twice performs a full $2\pi$ twist on the annulus; this is precisely a Dehn twist about the corresponding canonical curve.}
\label{fig:dehntriangles}
\end{figure}

We now show how to triangulate a mapping torus $T_{g, f}$ with $f = f_1f_2f_3\cdots f_n$ specified as a word in the standard generators. We begin with a triangulation of $\Sigma_g$ which has a strip of triangles along each of the canonical curves shown in Figure \ref{fig:dehn-twist}. This can be accomplished by choosing an appropriate triangulation of the genus $1$ surface with one deleted disk, and of the genus $2$ surface with two deleted disks, and then gluing $g$ of these together appropriately. To the resulting triangulation of $\Sigma_g$ we attach a shell of tetrahedra which implement $f_1$, that is, a triangulation of the mapping cylinder $M_{g, f_1}$. This shell is made up of tetrahedra which are always attached to the existing triangulation along two of their triangular faces, resulting in a 2-2 move; how to perform a Dehn twist using 2-2 moves is explained below, and illustrated in Figure \ref{fig:dehntriangles}. We repeat this procedure for each $f_i$ until we have a complete mapping cylinder $M_{g, f}$. To turn this into $T_{g, f}$, we need only glue the top triangles to the bottom ones. To turn the resulting triangulation into a tensor network, decorate each edge of $\Sigma_g$ with the gluing tensor $d_i$ (at an approximation scale cost of $O(2^g)$), and decorate each tetrahedron with the unitary 2-2 move tensor from Section \ref{sec:tet-unitary} (at an approximation scale cost of $1$). 

If instead we want to produce a triangulation of $H_{g, f}$ from a triangulation of $M_{g, f}$, we need to slightly modify the above procedure. First, we start with a handlebody triangulation (instead of just a triangulation of $\Sigma_g$) which also has a strip of triangles along each of the canonical curves on its surface. We then attach the triangulation of $M_{g, f}$ to the surface of the handlebody. We finish by taking another copy of the triangulated handlebody, and gluing the top of $M_{g, f}$ to its surface. We turn this triangulation into a tensor network just as in the mapping torus case, except we also add symbol tensors for each interior tetrahedron of the two handlebodies; the total approximation scale is still $O(2^g)$.

Finally, we specify how to implement a Dehn twist around some loop $\gamma$ on a triangulated surface. We can implement this twist (in either direction) by performing a particular sequence of 2-2 moves on the edges adjacent to $\gamma$, as illustrated in Figure \ref{fig:dehntriangles} (see, for instance,~\cite{BrunetNakamotoNegami} or \cite{KonigKuperbergReichardt}.) We assume that each canonical curve will lie along a strip of $2k$-many adjacent triangles for some integer $k$, as shown. By applying an ambient isotopy, we can make the two sides of the strip into concentric circles, with the vertices spaced out evenly. Order the vertices $\{v_i\}_{i=1}^k$ along $\gamma$ in a counterclockwise manner. Now do the following four times: for each $i$ from $1$ to $k$, flip the clockwisemost edge out of $v_i$ that does not lie on $\gamma$ (dashed edge in Figure \ref{fig:dehntriangles}). Notice that each flip amounts to shifting the inside end of that edge by $2\pi/k$ clockwise along $\gamma$, and shifting the outside end by $2\pi/k$ counterclockwise along the outer circle of the triangle strip. There are $2k$ edges, so $4k$ flips are needed to complete a full $2\pi$ twist. Notice that, combinatorially, the resulting triangulation is identical to the previous one; one can see this by assigning unique labels to all of the edges before the Dehn twist is applied and checking that the incidence matrix of the graph $($vertices, edges$)$ is unchanged. This fact allows us to compose twists and thus build mapping cylinders $M_{g, f}$ for arbitrary Dehn words $f$.

We have thus proved the following.

\begin{thm}\label{thm:large-triangulations}
There is a constant $C$ such that for every fixed integer $g>0$, there exist triangulated 3-manifolds with an arbitrarily large number of tetrahedra, whose Turaev-Viro invariant can be approximated by a quantum algorithm with an additive approximation scale no worse than $C^g$.
\end{thm}
Fixing a particular $g$, we could take the families 
$$
\{H_{g, f}: f \in \text{MCG}(g)\} \qquad \text{or} \qquad \{T_{g, f}: f \in \text{MCG}(g)\}~,
$$
where words are chosen arbitrarily and all manifolds are triangulated according to the above prescription. Of course, in general we have no guarantee that these families consist of pairwise nonhomeomorphic manifolds. For instance, $H_{g, f} \cong H_{g, f'}$ if $f$ and $f'$ belong to the same double coset of the so-called handlebody subgroup of MCG$(g)$, consisting of self-homeomorphisms of $\Sigma_g$ which extend to the identity on the handlebody of genus $g$. It's also easy to see that $T_{g, f} \cong T_{g, hfh^{-1}}$ for any $h \in \text{MCG}(g)$. However, there does exist an infinite family of pairwise non-homeomorphic 3-manifolds specified as Heegaard splittings with genus one. These are the so-called lens spaces~\cite{PrasolovSossinsky}. The fact that such a family can be found in the simplest case $g=1$ makes it plausible to conjecture that such families also exist for larger $g$. 

The constant $C$ in Theorem \ref{thm:large-triangulations} is essentially the worst-case norm of contracting the tensor network corresponding to a single triangulated ``handle'', $g$ of which can be attached together to form a genus-$g$ handlebody. We can give a bound for $C$ by using the fact that the operator norm of a tensor, regardless of how it is contracted, is always bounded above by its Hilbert-Schmidt norm (i.e., its $L^2$-norm as a vector.) The squared $L^2$-norms of the symbol tensor and the gluing tensor are
$$
\sum_{i,j,k,l,m,n \in I} 
\begin{vmatrix} i & j & k \\ l & m & n \end{vmatrix}^2 = 16 - 4 \sqrt{5}
\qquad \qquad
\sum_{i \in I} d_i^2 = \frac{5+\sqrt{5}}{2}~,
$$
respectively. We then have
$$
C \leq (16-4 \sqrt{5})^{t/2} \left(\frac{5+\sqrt{5}}{2}\right)^{3t} = (50(45+19 \sqrt{5}))^{t/2}~,
$$
where $t$ is the minimum number of tetrahedra necessary to triangulate a single handle. While a handle can be triangulated with only two tetrahedra, more tetrahedra may be necessary to ensure that the canonical curves from Figure \ref{fig:dehn-twist} all lie along edges of the triangulation.

\subsubsection{BQP-hardness of approximating Turaev-Viro}

We now briefly outline a proof that the problem of approximating the Turaev-Viro invariant of a triangulated 3-manifold is BQP-hard, in an appropriate sense. Our proof will be by reduction from the following standard BQP-hard\footnote{more precisely, PromiseBQP-hard} problem~\cite{AharonovArad}: given a quantum circuit $U$ of $n$ gates on $g$ qubits, decide in time poly$(n, g)$ if $|\langle 0^{\otimes g} | U | 0^{\otimes g} \rangle|^2 \leq 1/3$ or $|\langle 0^{\otimes g} | U | 0^{\otimes g}|^2 \rangle \geq 2/3$.

Suppose that we are given a quantum circuit $U$ consisting of $n$ gates acting on a total of $g$ qubits. We give an efficient procedure for translating such a circuit into a triangulated 3-manifold $M$ (equipped with an ordering of the tetrahedra) such that $\op{TV}(M)$ approximates $|\langle 0^{\otimes g} | U | 0^{\otimes g}|^2$. This translation proceeds in two stages. First, using previous results~\cite{AlagicJordanKonigReichardt} we can efficiently translate $U$ into a word $w$ of length poly$(n,g)$ in the Dehn twist generators of MCG$(g)$, with the promise that $\op{TV}(H_{g, w})$ is within $1/6$ of $|\langle 0^{\otimes g} | U | 0^{\otimes g}|^2.$ This procedure relies on an alternative definition of $\op{TV}(M)$ as the modulus squared of a matrix entry in a certain representation $\rho$ of MCG$(g)$, and the fact that this representation has dense image in the unitary group for all $g \geq 2$~\cite{FreedmanLarsenWang}. The procedure from~\cite{AlagicJordanKonigReichardt} then guarantees that $\rho(w)$ approximates $U$ on an appropriate subspace. In the second stage, we use the procedure of the previous section to turn $w$ into an efficient triangulation of the 3-manifold $H_{g, w}$, along with an ordering of the tetrahedra. It is important that we use a small triangulation of the handlebodies, so that the approximation scale factor from Theorem \ref{thm:tv-algorithm} satisfies $\Delta D^{-2|V|} \leq 1$. This guarantees that the result of contracting the tensor network of this triangulation is still within $1/6$ of $|\langle 0^{\otimes g} | U | 0^{\otimes g}|^2$.

The above proof can be viewed as translating quantum states on $g$ qubits into labelings of a triangulated surface with genus $g$ and translating local quantum gates into local retriangulations of the surface, such that these two translations are consistent. In fact, this idea is used in~\cite{KonigKuperbergReichardt} to provide a new quantum error correcting code based on the Turaev-Viro invariant and the associated Topological Quantum Field Theory.

\section{Discussion}

A reasonable criticism of the algorithm presented in the previous section is that the approximation scale is still exponential in the genus (even though the triangulation could grow unconstrained by the genus.) There are a number of reasons why it is unlikely that a significantly better scale can be achieved. First, the problem is BQP-hard with the stated scale. Second, in the Heegaard splitting and mapping torus versions of the problem~\cite{AlagicJordan, AlagicJordanKonigReichardt}, the approximation scales are also exponential in the genus. The same is true (with bridge number of the link presentation substituted for the genus) for the BQP-complete and DQC1-complete versions of the Jones Polynomial problem~\cite{AharonovArad, ShorJordan} as well as the Turaev-Viro problem for 3-manifolds specified by Dehn surgery~\cite{GarneroneMarzuoliRasetti}. A recent work of Kuperberg shows that for these types of problems, it is unlikely that there are quantum algorithms whose approximation scale is not exponentially large in some presentation-dependent quantity~\cite{Kuperberg}. We also remark that exact calculation of the Turaev-Viro invariant is NP-hard~\cite{KirbyMelvin}.

There are many important similarities between the various results mentioned above. First, both the Jones Polynomial and the Turaev-Viro invariant can be viewed as matrix entries (or traces, depending on the presentation of the underlying link or manifold) of a representation of a mapping class group of a surface. In the case of the Jones Polynomial, the surface in question is the $n$-punctured sphere; in the case of the Turaev-Viro invariant, it is the torus of genus $g$. From this point of view, these invariants are natural candidates for approximation on a quantum computer. In a sense, this point of view also makes them natural candidates for BQP-hardness: in both cases, the hardness results rely on certain properties of these representations, density in the unitary group being the most essential. 

We remark that the BQP results on the Turaev-Viro invariant could be unified by an efficient classical algorithm for translating (in any direction) between three presentation types: triangulation, Heegaard splitting, and Dehn surgery. In the previous section, we described one such efficient method which converts Heegaard splittings and mapping tori into triangulations. Similar methods may be possible for converting Dehn surgeries into triangulations as well. On the other hand, no efficient procedure for converting a triangulation into a Heegaard splitting or a Dehn surgery is known at this time~\cite{Thurston}.

It is natural to consider extensions of the above problems in higher dimensions. The Crane-Yetter invariant of triangulated 4-manifolds, for instance, is a state-sum invariant analogous to Turaev-Viro~\cite{CraneYetter}. The Turaev-Viro invariant is defined by assigning a 6j-symbol to each tetrahedron, with one index assigned to each edge; the Crane-Yetter invariant is defined by assigning a 15j-symbol to each pentachoron, with one index assigned to each edge and each triangular face. Like the two-dimensional TLFT invariant associated to a finite group (but unlike Turaev-Viro) Crane-Yetter is known to have a closed-form expression~\cite{CraneYetterKauffman} involving the Euler characteristic and the so-called \emph{signature} of the manifold. The Euler characteristic is trivial to calculate exactly from homology; the signature has a simple description in terms of the cohomology ring and is thus likely to have an efficient exact classical algorithm as well~\cite{KaibelPfetsch}.

\section{Acknowledgements}

G.A. is indebted to Alex Russell and Cris Moore for first telling us about the idea of approximating topological invariants by contracting tensor networks. We thank Stephen Jordan, Robert K\"onig and Dylan Thurston for useful conversations. G.A. acknowledges the support of NSERC, MITACS, and the U.S. ARO. E.B. thanks Ashwin Nayak and the University of Waterloo Combinatorics \& Optimization department for accepting him into the summer URA program during which his contributions were conducted. E.B. acknowledges the support of NSERC.

\bibliographystyle{plain}
\bibliography{euler}

\end{document}